\documentclass[12pt]{iopart}

\usepackage{iopams}  
\usepackage{graphicx}

\usepackage{cite}

\newcommand\beq{\begin{equation}}
\newcommand\eeq{\end{equation}}

\newcommand{\xv}{{\hat {\bf x}}}
\newcommand{\yv}{{\hat {\bf y}}}

\begin{document}

\title[Enhanced Faraday Rotation via Resonant Tunnelling]{Enhanced Faraday Rotation via Resonant Tunnelling in Tri-Layers Containing Magneto-Optical Metals}

\author{Massimo Moccia,$^1$ Giuseppe Castaldi,$^1$ Vincenzo Galdi,$^1$ Andrea Al\`u $^2$ and Nader Engheta $^3$}

\address{$^1$ Waves Group, Department of Engineering, University of Sannio, I-82100 Benevento, Italy\\
$^2$ Department of Electrical and Computer Engineering, The University of Texas at Austin, Austin, TX 78712, USA\\
$^3$ Department of Electrical and Systems Engineering, University of Pennsylvania, Philadelphia, PA 19104, USA}

\ead{vgaldi@unisannio.it}
\begin{abstract}
We study resonant tunnelling effects that can occur in tri-layer structures featuring a dielectric layer sandwiched between two magneto-optical-metal layers. We show that the resonance splitting associated with these phenomena can be exploited to enhance Faraday rotation at optical frequencies. Our results indicate that, in the presence of realistic loss levels, a tri-layer structure of sub-wavelength thickness is capable of yielding sensible ($\sim 10^o$) Faraday rotation with transmittance levels that are an order of magnitude larger than those attainable with a standalone slab of magneto-optical metal of same thickness.
\end{abstract}

%Uncomment for PACS numbers title message
% Keywords required only for MST, PB, PMB, PM, JOA, JOB? 
%\vspace{2pc}
%\noindent{\it Keywords}: Article preparation, IOP journals
% Uncomment for Submitted to journal title message
%\submitto{\JPA}
% Comment out if separate title page not required
\maketitle

%%%%%%%%%%%%%%%%%%%%%%%%%%%%%%%%%%%%%%%%%%%%%%%%%%%%%%%%%%%%%%%%%%
\section{Introduction}
%%%%%%%%%%%%%%%%%%%%%%%%%%%%%%%%%%%%%%%%%%%%%%%%%%%%%%%%%%%%%%%%%%
Faraday rotation effect (FRE), i.e., the rotation of the plane of polarization of a linearly-polarized electromagnetic wave propagating in a gyrotropic medium, is one of the most basic magneto-optical (MO) interactions \cite{Kotov:2010}, and finds important applications to realize nonreciprocal devices such as isolators and circulators. In essence, this effect relies on the {\em circular birefringence} of MO materials, for which
circularly-polarized (CP) waves of opposite handedness (in terms of which a linearly-polarized wave can be decomposed) travel with
different phase velocities.

At optical frequencies, typical MO materials (such as bismuth iron garnet \cite{Adachi:2000,Tepper:2003}) exhibit a rather weak gyrotropic response. Nonetheless, significant response enhancements can be obtained via artificial microstructures such as
magneto-photonic crystals in the form of multilayers combining dielectric and MO materials \cite{Inoue:1997,Inoue:1998if,Sakaguchi:1999kj,Steel:2000di,Kato:2003,Kato:2003a,Kahl:2004,Goto:2008,Goto:2009}, magneto-plasmonic devices \cite{Chin:2013bg,Davoyan:2013fr,Davoyan:2013dt}, or epsilon-near-zero metamaterials \cite{Davoyan:2013dq}.

On the other hand, MO metals such as Co$_6$Ag$_{94}$ \cite{Wang:1999} exhibit much stronger gyrotropic responses but, due to their {\em inherent opaqueness}, are limited by very low transmission levels. However, two recent studies \cite{Dong:2010ez,Dong:2011fy} have shown the possibility to overcome this limitation (i.e., to significantly enhance the transmission) by pairing a thin MO-metal layer with an all-dielectric photonic or magneto-photonic crystal. From the physical viewpoint, this approach relies on resonant tunnelling effects that can occur in heterostructures containing {\em epsilon-negative} (ENG) and {\em mu-negative} (MNG) media \cite{Fredkin:2002,Lakhtakia:2003,Alu:2003hb}. More specifically, at optical frequencies, MO metals naturally behave as circularly-birefringent ENG media, whereas an all-dielectric photonic crystal operating in the bandgap region can effectively mimic an MNG-type behaviour \cite{Guo:2006}.
Overall, taking into account realistic losses, multilayers containing a dozen layers (with total thickness of about 1.5 wavelengths) were shown to provide Faraday rotation angles of $\sim 15^o$, with transmittance levels $\sim 0.3$ that are about an order of magnitude higher than those pertaining to a standalone MO-metal slab \cite{Dong:2010ez,Dong:2011fy}.

In this paper, we show that similar effects and comparable performance can be obtained with a simpler and more compact structure. Our approach is inspired by a series of studies on tunnelling effects occurring in heterostructures combining ENG and {\em double-positive} (i.e., with positive permittivity and permeability) material layers  \cite{Zhou:2005dr,Hou:2005,Feng:2009jk,Oraizi:2010,Castaldi:2011kg,Castaldi:2011ka,Butler:2011,Cojocaru:2011tl,Liu:2012}. In particular, we consider a tri-layer structure of subwavelength thickness ($\sim 0.3$ wavelengths) combining MO-metal and dielectric layers.

The rest of the paper is organized as follows. In section \ref{Sec:Main}, we introduce the problem geometry and formulation, and outline the physical intuition as well as the main analytical derivations. In section \ref{Sec:Results}, we present some numerical results in order to illustrate the FRE enhancement and quantitatively assess the performance, taking also into account realistic loss and dispersion effects. Finally, in section \ref{Sec:Conclusions}, we provide some concluding remarks and hints for future research.

%%%%%%%%%%%%%%%%%%%%%%%%%%%%%%%%%%%%%%%%%%%%%%%%%%%%%%%%%%%%%%%%%%
\section{Tunnelling and Enhanced FRE in Tri-Layers Containing MO Metals}
%%%%%%%%%%%%%%%%%%%%%%%%%%%%%%%%%%%%%%%%%%%%%%%%%%%%%%%%%%%%%%%%%%
\label{Sec:Main}

%=================================================================
\subsection{Problem Geometry and Formulation}
%=================================================================
Referring to the geometry depicted in figure \ref{Figure1}, we consider a tri-layer structure (stacked along the $z$ axis, of infinite extent in the $x,y$ plane, and immersed in vacuum) composed of a dielectric layer of thickness $l_d$ and relative permittivity $\varepsilon_d$ sandwiched between two identical MO-metal layers of thickness $l_{MO}$. In the time-harmonic [$\exp(-i\omega t)$] regime, and in the presence of a static bias magnetization oriented along the $z$ direction, such material is characterized by a gyrotropic relative permittivity tensor \cite{Landau:1984}
\beq
{\underline {\underline \varepsilon}}_{MO}=\left[
\begin{array}{ccc}
\varepsilon_1 & -i\varepsilon_2 & 0\\
i\varepsilon_2 & \varepsilon_1 & 0\\
0 & 0 & \varepsilon_3
\end{array}
\right].
\label{eq:eps1}
\eeq
We assume that all materials are nonmagnetic and, for the moment, we neglect losses. At optical frequencies, the MO metal described by  (\ref{eq:eps1}) is characterized by
\beq
\varepsilon_1<0, ~~ 0<\varepsilon_2\ll |\varepsilon_1|,
\label{eq:eps12}
\eeq
with $\varepsilon_3$ being irrelevant for our purposes. We consider the so-called Faraday configuration \cite{Landau:1984}, by assuming a normally-incident linearly-polarized wave, with wavevector parallel to the static bias magnetization, and $x-$directed unit-amplitude electric field
\beq
{\bf E}^{(in)}\left(z\right)=\exp\left(i k z\right)\xv,
\label{eq:Ei}
\eeq
where $k=\omega/c=2\pi/\lambda$ denotes the vacuum wavenumber (and $c$ and $\lambda$ the corresponding wavespeed and wavelength, respectively), boldface symbols identify vector quantities, and the caret \^{} tags unit vectors. Our study is aimed at determining the conditions under which the impinging wave is significantly transmitted (with low reflection) through the structure, preserving its linearly-polarized character but with a rotation of an angle $\theta_F$ (Faraday rotation angle \cite{Palik:1970}) in the polarization plane (cf. figure \ref{Figure1}).

%=================================================================
\subsection{Modeling and Physical Intuition}
%=================================================================

In the assumed conditions, the wave propagation in the MO metal is most naturally described in terms of {\em evanescent} (forward and backward) 
right-handed and left-handed CP (RCP and LCP, respectively) eigenwaves with electric fields \cite{Landau:1984}
\numparts
\begin{eqnarray}
{\bf E}_{RCP}^{(FW)}(z)&=&\exp\left(-\alpha_{(+)} z\right)\left(\xv+i\yv\right),
\label{eq:eigen1}\\
{\bf E}_{RCP}^{(BW)}(z)&=&\exp\left(\alpha_{(-)} z\right)\left(\xv-i\yv\right),\\
{\bf E}_{LCP}^{(FW)}(z)&=&\exp\left(-\alpha_{(-)} z\right)\left(\xv-i\yv\right),\\
{\bf E}_{LCP}^{(BW)}(z)&=&\exp\left(\alpha_{(+)} z\right)\left(\xv+i\yv\right),
\label{eq:eigen4}
\end{eqnarray}
\endnumparts
where
\beq
\alpha_{(+)}=k\sqrt{-\varepsilon_1+\varepsilon_2},~~
\alpha_{(-)}=k\sqrt{-\varepsilon_1-\varepsilon_2},
\label{eq:alphapm}
\eeq
denote the [real valued, in view of (\ref{eq:eps12})] attenuation constants, which depend on the CP handedness and propagation direction. It is therefore expedient to decompose the linearly-polarized incident field in (\ref{eq:Ei}) in terms of RCP and LCP components, 
\beq
{\bf E}^{(in)}\left(z\right)={\bf E}^{(in)}_{RCP}\left(z\right)+{\bf E}^{(in)}_{LCP}\left(z\right),
\label{eq:Ei1}
\eeq
with
\beq
{\bf E}^{(in)}_{RCP,LCP}\left(z\right)=\exp\left(ikz\right)\left(\frac{\xv\pm i\yv}{2}\right),
\eeq
and invoke the superposition principle in order to reduce to a better problem-matched CP illumination. As anticipated, FRE is essentially based on the different phase velocities (and hence different phase accumulations) of RCP and LCP waves traveling in an MO material. However, given the evanescent character of the eigenwaves in (\ref{eq:eigen1})--(\ref{eq:eigen4}), it is fairly evident that a standalone MO-metal slab would be {\em essentially opaque}. To give an idea, assuming $\varepsilon_1=-10.51$ and $\varepsilon_2=1.15$ (realistic values, aside for losses, at $\lambda=631$nm \cite{Wang:1999}), a standalone MO-metal slab of thickness $0.1\lambda$ would yield a transmittance as low as $0.023$. 

From the studies in \cite{Alu:2003hb}, it is known that the opaqueness of an isotropic ENG slab can be ``compensated'' by pairing it with a suitably matched MNG slab. Although these results may be readily extended to our gyrotropic case here, synthesizing MNG materials at optical frequencies is a very challenging task from the technological viewpoint. As a possible surrogate, the use of an all-dielectric photonic crystal or a magneto-photonic crystal operating in the bandgap was suggested in \cite{Dong:2010ez} and \cite{Dong:2011fy}, respectively.

There are, however, other ways to compensate an ENG slab by means of simpler, more compact, and yet non-magnetic, structures, based, e.g., on the studies in \cite{Zhou:2005dr,Hou:2005,Feng:2009jk,Oraizi:2010,Castaldi:2011kg,Castaldi:2011ka,Butler:2011,Cojocaru:2011tl,Liu:2012}. The tri-layer structure in figure \ref{Figure1} is inspired by the isotropic ENG-dielectric-ENG tri-layer considered in \cite{Cojocaru:2011tl}, for which resonant tunnelling (with perfect transmission, in the absence of losses) was demonstrated. 
Extensions of such isotropic structure to the gyrotropic case were considered in \cite{Chen:2012et,Moccia:2013}, by substituting the central dielectric layer with an MO material, thereby achieving one-directional tunnelling of CP waves. Here, instead, we substitute the two ENG layers with MO-metal ones, and derive the resonant tunnelling conditions for RCP and LCP waves. Contrary to the studies in \cite{Chen:2012et,Moccia:2013}, we are not interested in achieving strong sensitivity of the tunnelling phenomena with respect to the CP handedness, but rather to attain high transmission of {\em both} RCP and LCP components in (\ref{eq:Ei1}), so as to preserve the linearly-polarized character of the impinging field.

%=================================================================
\subsection{Analytical Results and Tunnelling Conditions}
%=================================================================
Assuming an RCP or LCP excitation, the electromagnetic response of the tri-layer structure in figure \ref{Figure1} can be computed analytically in a rather straightforward fashion. Referring to \ref{Sec:APPA} for details, the transmission and reflection coefficients for RCP excitation ($T_{RCP}$ and $R_{RCP}$, respectively) can be written as
\numparts
\begin{eqnarray}
T_{RCP}&= &\frac{2ik k_d \alpha_{(+)}^2 \mbox{sech}\left(\alpha_{(+)}l_{MO}\right) \sec\left(k_dl_d\right)}{D_R},
\label{eq:TRCP}\\
D_R&=&
\tau_d\left[\alpha_{(+)}^2\left(\alpha_{(+)}\tau_{(+)}-ik\right)^2-k_d^2\left(\alpha_{(+)}-ik\tau_{(+)}\right)^2\right]\nonumber\\
&-&2 k_d\alpha_{(+)}\left(k\tau_{(+)}+i\alpha_{(+)}\right)\left(k+i\alpha_{(+)}\tau_{(+)}\right),
\end{eqnarray}
\endnumparts
\numparts
\begin{eqnarray}
R_{RCP}&=&\frac{N_{RR}}{D_R},
\label{eq:RRCP}\\
N_{RR}&=&\tau_d\left[\alpha_{(+)}^2\left(k_d^2-k^2\right)+\left(k^2k_d^2-\alpha_{(+)}^4\right)\tau_{(+)}^2\right]\nonumber\\
&-&2k_d\alpha_{(+)}\tau_{(+)}\left(k^2+\alpha_{(+)}^2\right),
\label{eq:NRR}
\end{eqnarray}
\endnumparts
where
\beq
k_d=k\sqrt{\varepsilon_d},~~\tau_d=\tan\left(k_dl_d\right),
\eeq
\beq
\tau_{(+)}=\tanh\left(\alpha_{(+)}l_{MO}\right).
\eeq
For the LCP case, results are formally analogous, by replacing $\alpha_{(+)}$ with $\alpha_{(-)}$, viz.,
\numparts
\begin{eqnarray}
T_{LCP}&=&\frac{2ik k_d \alpha_{(-)}^2 \mbox{sech}\left(\alpha_{(-)}l_{MO}\right)\sec\left(k_dl_d\right)}{D_L},\label{eq:TLCP}\\
D_L&=&\tau_d\left[\alpha_{(-)}^2\left(\alpha_{(-)}\tau_{(-)}-ik\right)^2-k_d^2\left(\alpha_{(-)}-ik\tau_{(-)}\right)^2\right]\nonumber\\
&-&2 k_d\alpha_{(-)}\left(k\tau_{(-)}+i\alpha_{(-)}\right)\left(k+i\alpha_{(-)}\tau_{(-)}\right),
\end{eqnarray}
\endnumparts
\numparts
\begin{eqnarray}
R_{LCP}&=&\frac{N_{RL}}{D_L},
\label{eq:RLCP}\\
N_{RL}&=&\tau_d\left[\alpha_{(-)}^2\left(k_d^2-k^2\right)+\left(k^2k_d^2-\alpha_{(-)}^4\right)\tau_{(-)}^2\right]\nonumber\\
&-&2k_d\alpha_{(-)}\tau_{(-)}\left(k^2+\alpha_{(-)}^2\right),
\label{eq:NRL}
\end{eqnarray}
\endnumparts
with
\beq
\tau_{(-)}=\tanh\left(\alpha_{(-)}l_{MO}\right).
\eeq
From (\ref{eq:RRCP}) and (\ref{eq:RLCP}), the conditions for zero reflection (i.e., total transmission) for RCP and LCP excitation are readily derived as
\beq
l_d=\frac{1}{k_d}\arctan\left[\frac{2k_d\alpha_{(\pm)}\tau_{(\pm)}\left(k^2+\alpha_{(\pm)}^2\right)}{\alpha_{(\pm)}^2\left(k_d^2-k^2\right)+\left(k^2k_d^2-\alpha_{(\pm)}^4\right)\tau_{(\pm)}^2}\right],
\label{eq:ld}
\eeq
where the $(+)$ and $(-)$ subscripts apply to RCP and LCP, respectively. In other words, for a given CP handedness, and assigned frequency and MO-metal layer parameters ($l_{MO}$, $\varepsilon_1$, $\varepsilon_2$), it is {\em always} possible to find a dielectric layer of {\em arbitrary} relative permittivity $\varepsilon_d$ and thickness $l_d$ chosen according to (\ref{eq:ld}) (plus its infinite half-wavelength periodicities) so that the power effectively tunnels through the tri-layer and is totally transmitted, despite the inherent opaqueness of the MO-metal constituents. We highlight that the guaranteed existence of a total-transmission solution as well as the completely arbitrary choice of the dielectric material make the tri-layer structure in figure \ref{Figure1} preferable to alternative tri-layer configurations such as, e.g., that in \cite{Zhou:2005dr} (featuring an ENG layer sandwiched between two dielectric layers), or that in \cite{Castaldi:2011kg} (featuring an ENG layer paired at one side with a dielectric bi-layer).

It is clear from (\ref{eq:ld}) that, as a consequence of the aforementioned circular birefringence, the tunnelling conditions for the RCP and LCP excitations are {\em different}. Such {\em resonance-splitting} phenomenon is illustrated in figure \ref{Figure2}, which, for realistic parameters (aside for losses) of the MO-metal layer, shows the two transmittances as a function of the dielectric layer ($\varepsilon_d=2.12$) electrical thickness. As it can be observed, the responses are characterized by a series of full-transmission peaks that are slightly shifted in the RCP and LCP cases. Therefore, in order to preserve the linearly-polarized character of the impinging field in (\ref{eq:Ei1}), we cannot work at either of the RCP or LCP full-transmission peaks, but we rather need to settle to a {\em midway} condition for which
\beq
\left|T_{RCP}\right|=\left|T_{LCP}\right|.
\label{eq:crossing}
\eeq
For the assumed parameter configuration, this choice still allows a significantly high transmittance of $\sim 0.8$ [see the magnified details in figure \ref{Figure2}(b)], i.e., $\sim 35$ times higher than that attainable in the presence of a standalone MO-metal slab of thickness $2l_{MO}=0.1\lambda$. There is an evident trade-off between the magnitude of attainable Faraday rotation, which benefits from a larger distance between transmission peaks for LCP and RCP waves, and the amount of total transmission level to achieve linearly-polarized output, which instead grows for smaller shifts.

%%%%%%%%%%%%%%%%%%%%%%%%%%%%%%%%%%%%%%%%%%%%%%%%%%%%%%%%%%%%%%%%%%
\section{Representative Numerical Results}
%%%%%%%%%%%%%%%%%%%%%%%%%%%%%%%%%%%%%%%%%%%%%%%%%%%%%%%%%%%%%%%%%%
\label{Sec:Results}

%=================================================================
\subsection{Generalities and Observables}
%=================================================================
The observables of interest in our study can be computed from the transmission and reflection coefficients in (\ref{eq:TRCP})--(\ref{eq:NRL}). In what follows, assuming the linearly-polarized excitation in (\ref{eq:Ei}), 
we compute the total transmittance [cf. (\ref{eq:Ei1})] as
\beq
\left|T\right|^2=\frac{1}{2}\left(\left|T_{RCP}\right|^2+\left|T_{LCP}\right|^2\right).
\label{eq:TT}
\eeq
Moreover, for quantitative assessment of the FRE enhancement, we compute the Faraday rotation angle \cite{Palik:1970}
\beq
\theta_F=\frac{1}{2}\arctan\left(\frac{T_{RCP}}{T_{LCP}}\right).
\label{eq:thetaF}
\eeq
Finally, to quantify the deviations from linear-polarization in the transmitted field, we compute the Faraday ellipticity \cite{Palik:1970}
\beq
\delta_F=\frac{\left|T_{RCP}\right|-\left|T_{LCP}\right|}{\left|T_{RCP}\right|+\left|T_{LCP}\right|}.
\label{eq:deltaF}
\eeq
In the more realistic lossy case (cf. section \ref{Sect:Lossy} below), we also compute the total reflectance
\beq
\left|R\right|^2=\frac{1}{2}\left(\left|R_{RCP}\right|^2+\left|R_{LCP}\right|^2\right).
\label{eq:RR}
\eeq

%=================================================================
\subsection{Ideal Lossless Case}
%=================================================================
We begin considering the ideal lossless case, by assuming the geometry and parameters as in figure \ref{Figure2}.
Figure \ref{Figure3} shows the three observables in (\ref{eq:TT})--(\ref{eq:deltaF}) as a function of the dielectric layer electrical thickness. We can observe a series of equispaced, identical resonances characterized by high-transmittance peaks [figure \ref{Figure3}(a)], accompanied by peaks in the Faraday rotation angle [figure \ref{Figure3}(b)], and abrupt variations (with zero crossing) of the Faraday ellipticity [figure \ref{Figure3}(c)]. 

Next, we set the dielectric layer thickness to $l_d=0.234\lambda$, so that the crossing condition in (\ref{eq:crossing}) is met at the operational wavelength [cf. figure \ref{Figure2}(b)]. This implies that, at that specific wavelength, the transmitted field will be linearly polarized. For instance, assuming (as in \cite{Dong:2010ez,Dong:2011fy}) a nominal operational wavelength $\lambda=631$ nm (i.e., $l_d=147$nm), figure \ref{Figure4} shows the arising observables, as a function of the wavelength. It can be observed that, at the nominal operational wavelength, the Faraday ellipticity vanishes, thereby confirming the linear-polarization character of the transmitted field. This is accompanied by high transmittance ($\sim 0.8$) and significant Faraday rotation angle ($\sim -28^o$), close to their peak values. To better understand the enhancement effects, it is worth recalling that a standalone MO-metal slab of thickness $2l_{MO}=0.1\lambda$ would yield a transmittance $|T|^2=0.023$, a Faraday rotation angle $\theta_F=-2.02^o$, and a Faraday ellipticity $\delta_F=-0.157$.

Figure \ref{Figure5} shows the electric (solid) and magnetic (dashed) field intensity distributions along the $z$ axis, inside and outside the tri-layer structure, at the nominal operational wavelength. Besides the high-transmission and low-reflection values already observable from figure \ref{Figure4}, we notice a sensible field enhancement inside the tri-layer, 
with electric and magnetic fields peaked at the center of the dielectric layer and at its boundaries with the MO-metal layers, respectively, accompanied by evanescent field growth in the left MO-metal layer. We note that the response resembles that of a Fabry-Perot resonant cavity, with the end layers providing both the required high reflectivity and the MO activity.

%=================================================================
\subsection{Realistic Lossy Case}
%=================================================================
\label{Sect:Lossy}
For a more realistic performance assessment, we need to take into account the inevitable loss and dispersion effects in the MO-metal. Figure \ref{Figure6} shows the dispersion laws for the real and imaginary parts of the relevant constitutive parameters $\varepsilon_1$ and $\varepsilon_2$ in (\ref{eq:eps1}) obtained by digitizing the experimental results in \cite{Wang:1999}, pertaining to Co$_6$Ag$_{94}$ measured at room temperature after annealing at $250^o$ C. For the dielectric layer, assuming fused silica (SiO$_2$) as a material, a lossless, nondispersive model with $\varepsilon_d=2.12$ can be safely adopted within the wavelength range of interest \cite{Malitson:1965}.

Figure \ref{Figure7} shows the same observables as in figure \ref{Figure4}, but assuming now the MO-metal dispersion laws of figure \ref{Figure6}. In addition, the reflectance $|R|^2$ in (\ref{eq:RR}) is also shown (blue-dashed). Moreover, as a reference, also shown (red-dotted) are the corresponding observables pertaining to a standalone MO-metal slab of thickness $2d_{MO}=64$nm. By comparison with the ideal lossless case in  figure \ref{Figure4}, at the nominal operational wavelength $\lambda=631$nm, we note a sensible reduction of the transmittance ($|T|^2\approx 0.3$, with reflectance $|R|^2\approx0.3$), which is, however, still over an order of magnitude higher that that achievable with a standalone MO-metal slab. Moreover, though reduced, the Faraday rotation angle ($\theta_F\approx -12^o$) is still enhanced by nearly a factor of three in comparison to the standalone MO-metal slab. Overall, the above results are in line with what observed in \cite{Dong:2010ez,Dong:2011fy}, but our configuration is structurally simpler and more compact, its total thickness being only $\sim0.3\lambda$ at the nominal operational wavelength. On the other hand, the Faraday ellipticity ($\delta_F\approx-0.35$), not considered in \cite{Dong:2010ez,Dong:2011fy}, is non-negligible.

It should be pointed out that our design procedure is based on the crossing condition of the split RCP and LCP resonances [cf. (\ref{eq:crossing})] in the ideal lossless case. However, the MO-metal losses affect the two CP polarizations in different ways. Therefore, in principle, better tradeoff configurations of the relevant observables may be found  by taking into account the loss effects at the design stage. As an example, the contour plots in figure \ref{Figure8} show the three observables in (\ref{eq:TT})--(\ref{eq:deltaF}), at the nominal operational wavelength, as a function of the MO-metal and dielectric layer thicknesses ($l_{MO}$ and $l_d$, respectively) when MO-metal losses (cf. figure \ref{Figure6}) are taken into account. From these plots, one may extract various tradeoff curves. For instance, for a targeted transmittance level, one may extract from figure \ref{Figure8}(a) the iso-transmittance ($l_{MO},l_d$) pairs, and derive from figures \ref{Figure8}(b) and \ref{Figure8}(c) a Faraday rotation angle vs. ellipticity tradeoff curve. Figure \ref{Figure9} shows three of such tradeoff curves (for targeted transmittance levels $|T|^2=0.2, 0.3$, and $0.4$) where each point corresponds to a possible combination of the layer thicknesses $l_{MO}$ and $l_d$ extracted from figure \ref{Figure8}.
It can be observed that, by comparison with figure \ref{Figure7}, tradeoff choices exist that, while maintaining comparable levels of transmittance ($|T|^2\sim 0.3$) and Faraday rotation angle ($\theta_F\sim -10^o$), yield sensibly smaller ellipticities $\delta_F\sim-0.14$.

%%%%%%%%%%%%%%%%%%%%%%%%%%%%%%%%%%%%%%%%%%%%%%%%%%%%%%%%%%%%%%%%%%
\section{Conclusions and Perspectives}
%%%%%%%%%%%%%%%%%%%%%%%%%%%%%%%%%%%%%%%%%%%%%%%%%%%%%%%%%%%%%%%%%%
\label{Sec:Conclusions}
In this paper, we have shown that enhanced FRE can be induced by resonant tunnelling in tri-layer structures featuring a dielectric layer sandwiched between MO-metal layers. Our analytical results allow for the derivation of simple design formulas, and provide a clear physical explanation of the phenomenon. In addition, we have assessed the performance in the presence of realistic loss and dispersion effects. 

Overall, by comparison with alternative multilayered configurations \cite{Dong:2010ez,Dong:2011fy}, we obtain comparable performance with a structurally simpler and thinner ($\sim 0.3\lambda$) configuration.

As possible follow-up studies, we plan to explore other nonreciprocal tunnelling phenomena inspired by the configurations in \cite{Zhou:2005dr,Hou:2005,Feng:2009jk,Oraizi:2010,Castaldi:2011kg,Castaldi:2011ka,Butler:2011,Cojocaru:2011tl,Liu:2012}. Also of interest are approaches to mitigating the effects of losses in the MO material constituents, e.g., along the lines of \cite{Figotin:2008hr}.

\let\thefigureSAVED\thefigure
\appendix
\let\thefigure\thefigureSAVED 

%%%%%%%%%%%%%%%%%%%%%%%%%%%%%%%%%%%%%%%%%%%%%%%%%%%%%%%%%%%%%%%%%%
\section{Derivation of (\ref{eq:TRCP})--(\ref{eq:NRL})}
\begin{figure}
\begin{center}
\includegraphics [width=10cm]{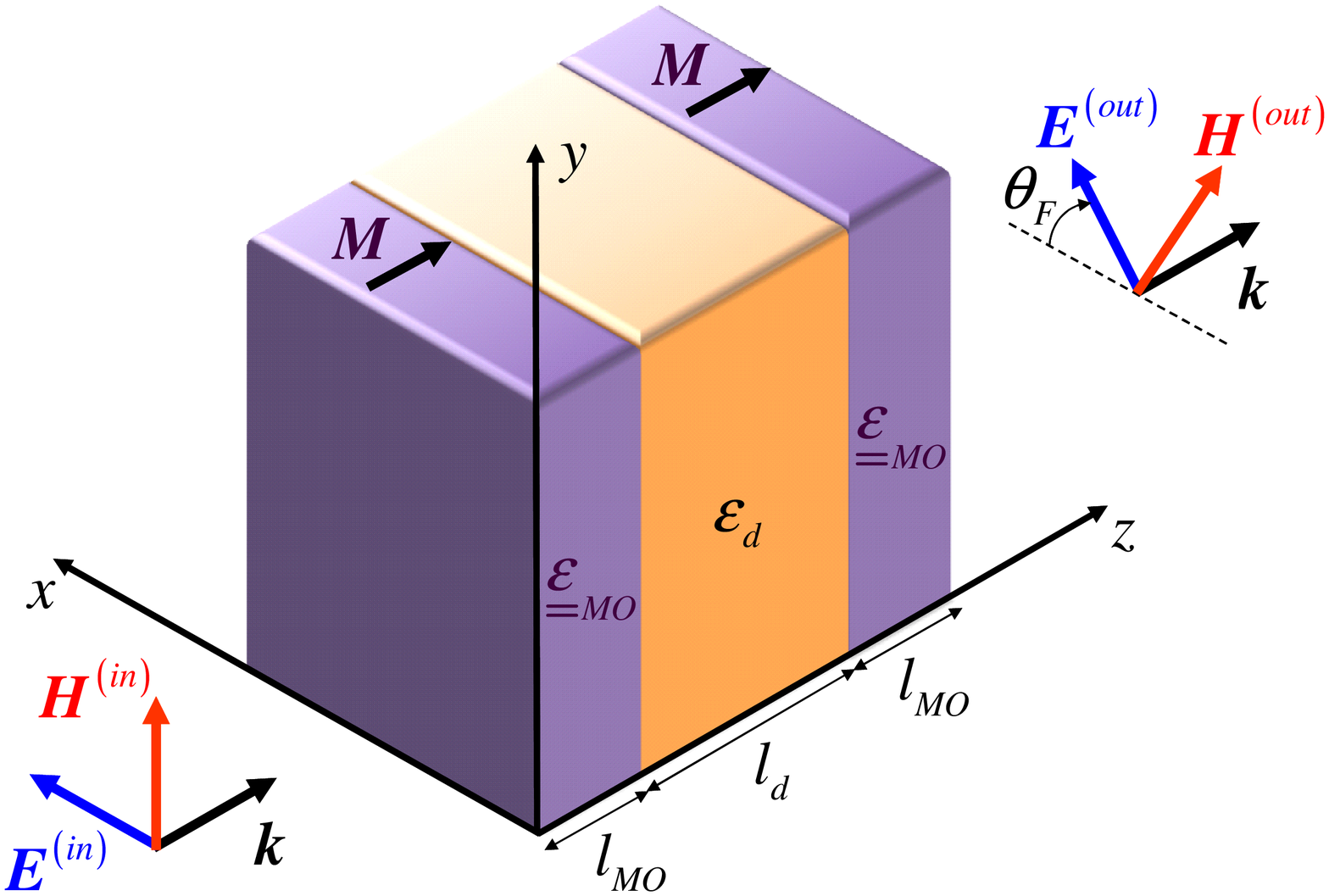}% Here is how to import EPS art
\end{center}
\caption{(Color online) Problem schematic: A tri-layer structure (stacked along the $z$ direction, of infinite extent in the $x,y$ plane, and immersed in vacuum) composed of a dielectric layer of thickness $l_d$ and relative permittivity $\varepsilon_d$ sandwiched between two identical layers of MO-metal layers of thickness $l_{MO}$ and relative permittivity tensor ${\underline {\underline \varepsilon}}_{MO}$ [assuming a static bias magnetization along the $z$-direction, cf. (\ref{eq:eps1})]. The structure is illuminated by a time-harmonic, linearly-polarized, normally-incident plane wave, with $x-$directed electric field. Ideally, the transmitted field is still linearly polarized, but with a rotation of an angle $\theta_F$ (Faraday rotation angle) of the plane of polarization.}
\label{Figure1}
\end{figure}

\begin{figure}
\begin{center}
\includegraphics [width=10cm]{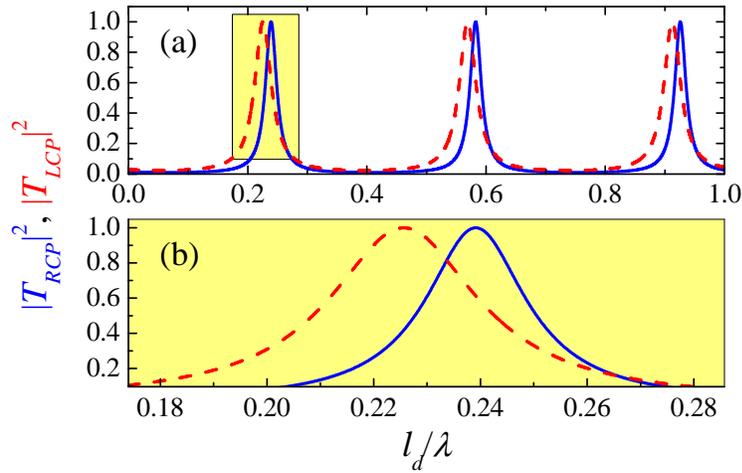}% Here is how to import EPS art
\end{center}
\caption{(Color online) Geometry as in figure \ref{Figure1}, with $\varepsilon_1=-10.51$, $\varepsilon_2=1.15$, $l_{MO}=0.05\lambda$, and $\varepsilon_d=2.12$. (a) Transmittances of CP waves, $|T_{RCP}|^2$ [cf. (\ref{eq:TRCP}), blue-solid curves] and $|T_{LCP}|^2$ [cf. (\ref{eq:TLCP}), red-dashed curves], as a function of the dielectric layer electrical thickness. (b) Magnified detail around the lowest-order resonances, highlighting the resonance splitting.}
\label{Figure2}
\end{figure}

\begin{figure}
\begin{center}
\includegraphics [width=10cm]{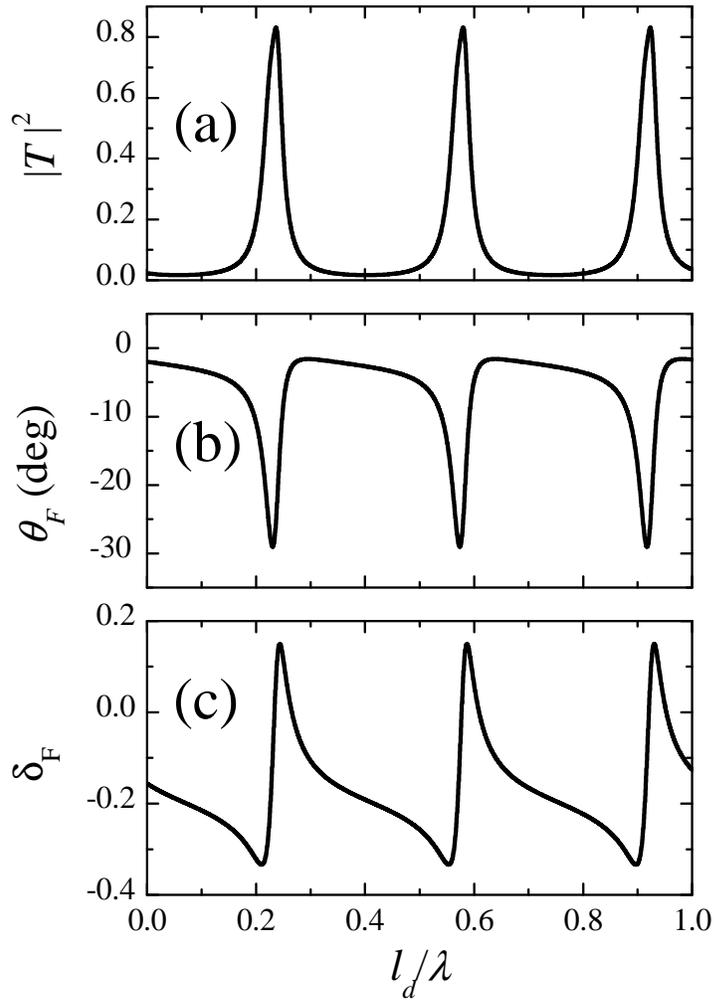}% Here is how to import EPS art
\end{center}
\caption{Parameters as in figure \ref{Figure2}. (a) Total transmittance [cf. (\ref{eq:TT})], (b) Faraday rotation angle [cf. (\ref{eq:thetaF})], (c) Faraday ellipticity [cf. (\ref{eq:deltaF})], as a function of the dielectric layer electrical thickness.}
\label{Figure3}
\end{figure}

\begin{figure}
\begin{center}
\includegraphics [width=10cm]{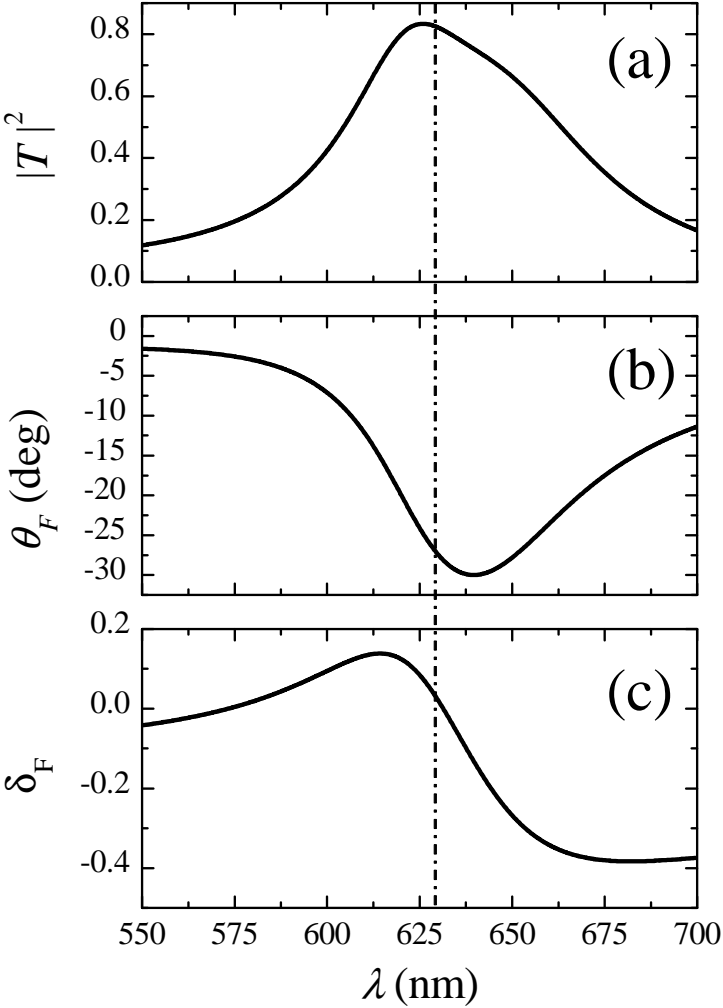}% Here is how to import EPS art
\end{center}
\caption{As in figure \ref{Figure3}, but as a function of wavelength, assuming $l_{MO}=32$nm and  $l_d=147$nm. The vertical dotted-dashed line indicates the nominal operational wavelength $\lambda=631$nm.}
\label{Figure4}
\end{figure}

\begin{figure}
\begin{center}
\includegraphics [width=10cm]{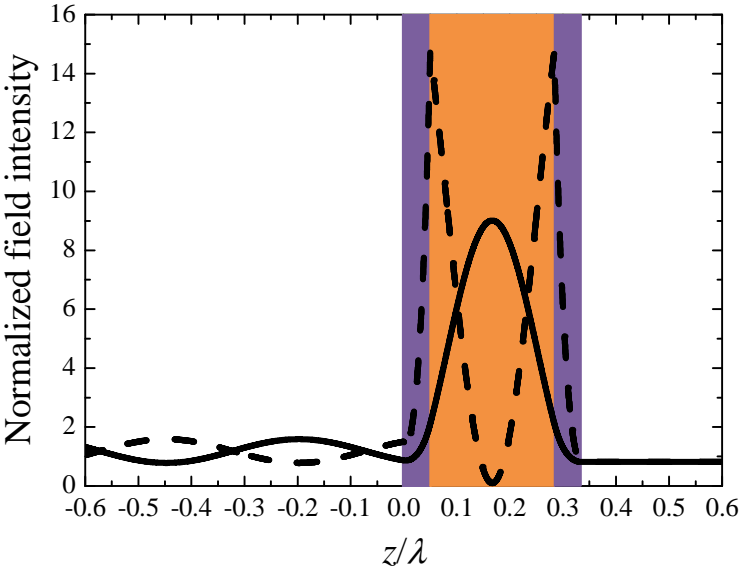}% Here is how to import EPS art
\end{center}
\caption{(Color online) Parameters as in figure \ref{Figure4}. Electric (solid) and magnetic (dashed) field-intensity distributions along the $z$ axis [normalized by the incident field intensities, cf. (\ref{eq:Ei})] at the nominal operational wavelength ($\lambda=631$nm). Different colors (consistent with figure \ref{Figure1}) are used in order to visually delimit the various material layers.}
\label{Figure5}
\end{figure}

\begin{figure}
\begin{center}
\includegraphics [width=10cm]{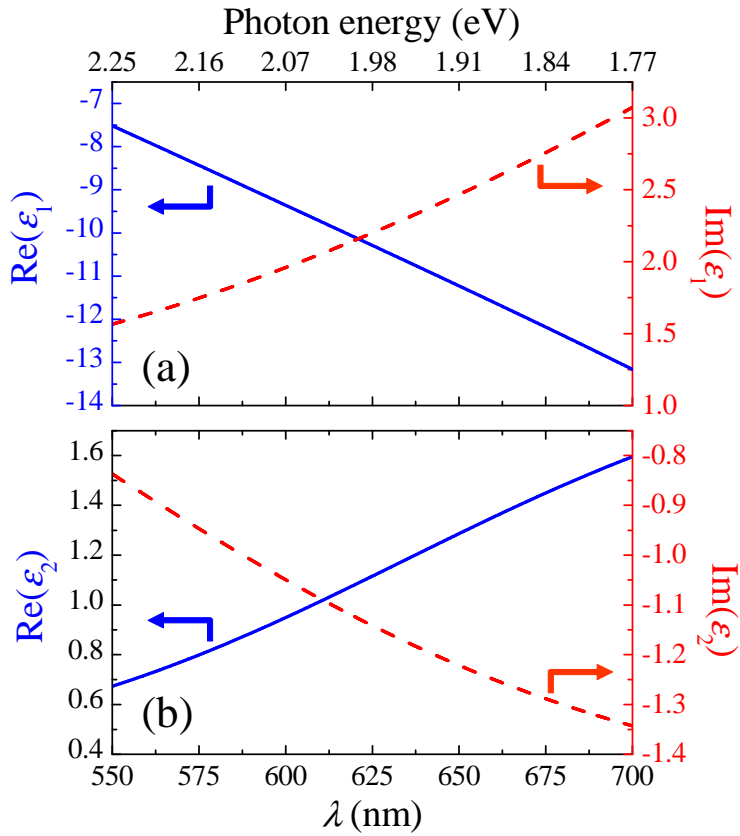}% Here is how to import EPS art
\end{center}
\caption{(Color online) (a), (b) Dispersion laws for the real (blue-solid curves) and imaginary (red-dashed curves) parts of the relevant MO-metal relative permittivity components $\varepsilon_1$ and $\varepsilon_2$, respectively, in (\ref{eq:eps1}). The curves are obtained by digitizing the experimental results in \cite{Wang:1999}, pertaining to Co$_6$Ag$_{94}$ measured at room temperature after annealing at $250^o$ C.}
\label{Figure6}
\end{figure}

\begin{figure}
\begin{center}
\includegraphics [width=10cm]{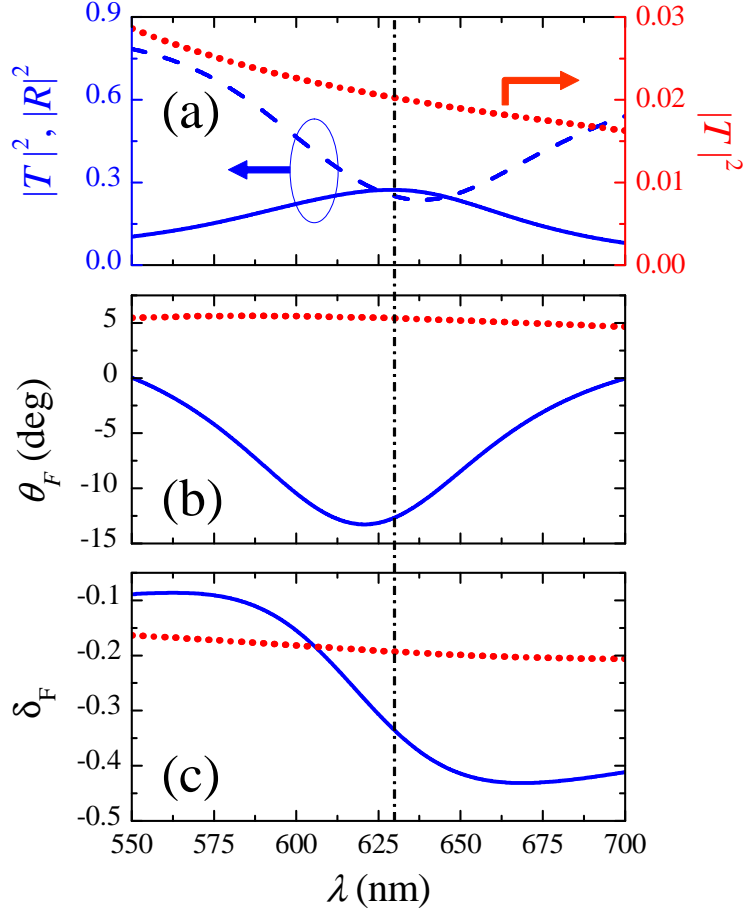}% Here is how to import EPS art
\end{center}
\caption{(Color online) As in figure \ref{Figure5}, but assuming the MO-metal dispersion laws in figure \ref{Figure6}, and a lossless, nondispersive fused-silica (SiO$_2$) dielectric layer ($\varepsilon_d=2.12$). Also shown are the reflectance $|R|^2$ in (\ref{eq:RR}) (blue-dashed curve) and the observables (red-dotted curves) pertaining to a standalone MO-metal slab of thickness $2l_{MO}=64$nm.}
\label{Figure7}
\end{figure}

\begin{figure*}
\begin{center}
\includegraphics [width=16cm]{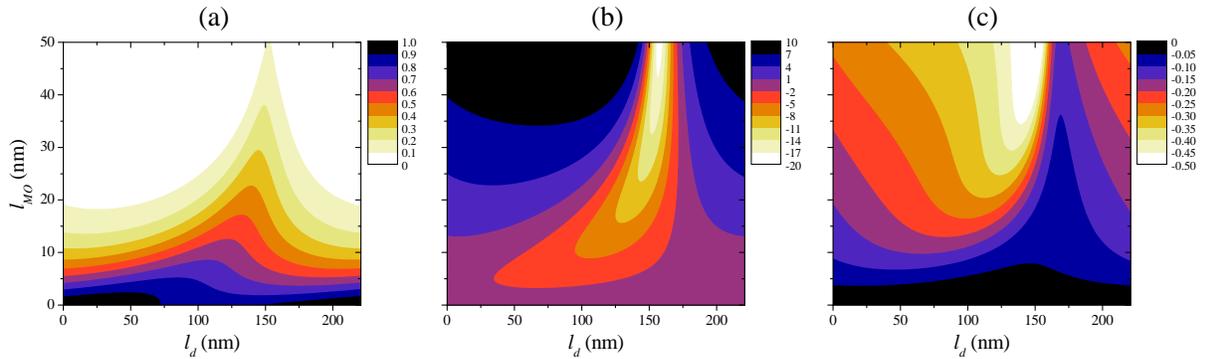}% Here is how to import EPS art
\end{center}
\caption{(Color online) (a), (b), (c) Transmittance $|T|^2$, Faraday rotation angle $\theta_F$ (deg), and ellipticity $\delta_F$, respectively, at the nominal design wavelength $\lambda=631$nm, as a function of the MO-metal and dielectric layer thicknesses. All other parameters are as in figure \ref{Figure7}.}
\label{Figure8}
\end{figure*}

\begin{figure}
\begin{center}
\includegraphics [width=10cm]{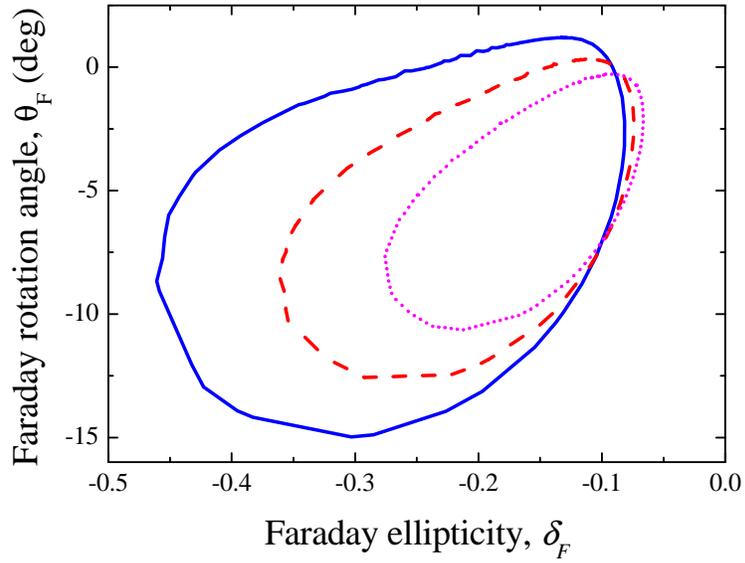}% Here is how to import EPS art
\end{center}
\caption{(Color online) Tradeoff curves showing the Faraday rotation angle $\theta_F$ vs. the ellipticity $\delta_F$, at the nominal operational wavelength, for targeted transmittance levels $|T|^2=0.2, 0.3$ and $0.4$ (blue-solid, red-dashed, and magenta-dotted curves, respectively). The curves are obtained by extracting the iso-transmittance ($l_{MO}$, $l_d$) pairs from figure \ref{Figure8}(a), and the corresponding $\theta_F$ and $\delta_F$ from figures \ref{Figure8}(b) and \ref{Figure8}(c).}
\label{Figure9}
\end{figure}

%%%%%%%%%%%%%%%%%%%%%%%%%%%%%%%%%%%%%%%%%%%%%%%%%%%%%%%%%%%%%%%%%%
\label{Sec:APPA}
Assuming a forward-propagating RCP excitation, we can express the electric field in the various regions of the structure in figure \ref{Figure1} as linear combinations of forward-propagating RCP waves and backward-propagating LCP waves, which are evanescent in the MO-metal layers, viz.,
\beq
{\bf E}\left(z\right)=e\left(z\right) \frac{\left(\xv+i\yv\right)}{2},
\label{eq:EE}
\eeq
with
\begin{eqnarray}
e\left(z\right)&=&
\exp\left(i k z\right)+B_0\exp\left(-i k z\right),~~z<0,
\label{eq:ee0}\\
e\left(z\right)&=&A_1 \exp\left(-\alpha_{(+)} z\right) +B_1\exp\left(\alpha_{(+)} z\right),\nonumber\\
&&0<z<l_{MO},
\label{eq:ee1}
\\
e\left(z\right)&=&A_2 \exp\left(i k_d z\right) +B_2\exp\left(-i k_d z\right),\nonumber\\ 
&&l_{MO}<z<l_{MO}+l_d,\\
e\left(z\right)&=&A_3 \exp\left(-\alpha_{(+)} z\right) +B_3\exp\left(\alpha_{(+)} z\right),\nonumber\\
&&l_{MO}+l_d<z<2l_{MO}+l_d,
\label{eq:ee2}\\
e\left(z\right)&=&A_4 \exp\left[ik \left(z-2l_{MO}-l_d\right)\right],\nonumber\\
&&z>2l_{MO}+l_d.
\label{eq:ee3}
\end{eqnarray}
The corresponding magnetic field can readily by derived from  (\ref{eq:EE})--(\ref{eq:ee3}) via the relevant Maxwell's curl equation
\beq
{\bf H}\left(z\right)=\frac{1}{i\omega\mu_0}\nabla \times {\bf E}\left(z\right),
\eeq
with $\mu_0$ denoting the vacuum permeability. Next, by enforcing the electric- and magnetic-field  continuity at the four interfaces $z=0$, $z=l_{MO}$, $z=l_{MO}+l_d$, and $z=2l_{MO}+l_d$, we obtain an $8\times 8$ linear system of equations whose (cumbersome but straightforward) solution yields the eight unknown coefficients $A_1$, $A_2$, $A_3$, $A_4$, $B_0$, $B_1$, $B_2$, and $B_3$ in (\ref{eq:ee0})--(\ref{eq:ee3}). In particular, the transmission and reflection coefficients given in (\ref{eq:TRCP})--(\ref{eq:NRR}) are directly given by the coefficients $A_4$ and $B_0$.

For the case of forward-propagating LCP excitation, the above procedure can be repeated in a formally analogous fashion, by assuming
\beq
{\bf E}\left(z\right)=e\left(z\right) \frac{\left(\xv-i\yv\right)}{2},
\label{eq:EE2}
\eeq
and replacing of $\alpha_{(+)}$ with $\alpha_{(-)}$ in (\ref{eq:ee1}) and (\ref{eq:ee2}). This yields the transmission and reflection coefficients in (\ref{eq:TLCP})--(\ref{eq:NRL}).

%\begin{thebibliography}{99}
%% Do not include separate BibTeX files; if BibTeX is used,
%% paste the output (contents of .bbl file) here.

%\bibliography{EFR_bibliography}
%\bibliographystyle{unsrt}

%\end{thebibliography}

%############################################################
%                Figure1
%
%############################################################

%############################################################
%                Figure2
%
%############################################################

%############################################################
%                Figure3
%
%############################################################

%############################################################
%                Figure4
%
%############################################################

%############################################################
%                Figure5
%
%############################################################

%############################################################
%                Figure6
%
%############################################################

%############################################################
%                Figure7
%
%############################################################

%############################################################
%                Figure8
%
%############################################################

%############################################################
%                Figure9
%
%############################################################

\end{document}